\documentstyle[11pt,paspconf,epsf]{article}

\begin{document}

\title{Red Companions to a z=2.15 Radio Loud Quasar}
\author{D.L. Clements}
\affil{Dept. Physics and Astronomy, Cardiff university, PO Box 913, Cardiff
CF22 3YB, UK\\
Institut d'Astrophysique Spatiale, Batiment 121, Univ. Paris XI, F-91405 ORSAY
Cedex, France\\
European Southern Observatory, Karl-Schwarzschild-Strasse 2, D-87548
Garching-bei-Munchen, Germany}

\begin{abstract}
We have conducted observations of the environment around the z=2.15
radio loud quasar 1550-269 in search of distant galaxies associated
either with it or the z=2.09 CIV absorber along its line of
sight. Such objects will be distinguished by their red colours,
R-K$>$4.5. We find five such objects in a 1.5 arcmin$^2$ field around the
quasar, with typical K magnitudes of $\sim$20.4 and no detected R band
emission. We also find a sixth object with K=19.6$\pm$0.3, and
undetected at R, just two arcseconds from the quasar. The nature of all these
objects is currently unclear, and will remain so until we have determined
their redshifts. We suggest that it is likely that they are associated with
either the quasar or the CIV absorber, in which case their properties
might be similar to those of the z=2.38 red Ly$\alpha$ emitting galaxies
discovered by Francis et al. (1997). The small separation between
the quasar and the brightest of our objects suggests that it may be the
galaxy responsible for the CIV metal line absorption system. The closeness
to the quasar and the red colour might have precluded similar objects from
being uncovered in previous searches for emission from CIV
and eg. damped absorbers.
\end{abstract}


\section{Introduction}

The selection of high redshift galaxies on the basis of their colours
has been a growth industry over the last 5 years, as demonstrated by
the breadth and depth of contributions to this meeting. Much of this
work has concentrated on the selection of high (z$>$3) redshift
objects through `dropout' techniques. Such methods have had
considerable success (see eg. C.C. Steidel's contribution in this
volume, among many others). However, many of these techniques are
reliant on emission in the rest-frame ultraviolet. The UV emission
from a galaxy can easily be dominated by a small burst of star
formation, or alternatively obscured by a relatively small amount of
dust. A population of older quiescent galaxies might thus coexist with
the UV selected high redshift objects. Studies of the stellar
populations in moderate redshift radio galaxies provide some support
for this idea.  A number of authors (eg. Stockton et al. (1995), Spinrad
et al. (1997)) have shown that several radio galaxies have ages
$>$3--5 Gyr at z$\sim$1.5, indicating that they must have formed at
z$>$5. These results have even been used (Dunlop et al. 1996) to argue that
$\Omega$ must be significantly less than 1.

Old galaxies at moderate redshift, passively evolving from z$>$5 to
z=1.5 -- 2.5, would appear as red objects, with R-K colours
$>$4.5. There has been considerable interest in such red objects. Much
of this work has centred on red objects found in the fields of known
high redshift AGN (eg. Hu \& Ridgeway (1994), Yamada et al.,
(1997)). A large survey of the environments of z=1--2 quasars (Hall et
al., 1998) finds that such associations are quite common. The present
paper attempts to push such studies above z=2. The alternative
approach, to study red objects in the field, is also an active area
with several surveys dedicated to or capable of finding such objects.
See eg. Cohen et al. (1998), or Rigopoulou et al. (in preparation).
There are also several other contributions in the present proceedings
on the subject.

We assume $\Omega_M$ = 1, $\Lambda$=0 and H$_0$=100 kms$^{-1}$Mpc$^{-1}$
throughout this paper.

\section{Observations}

As part of a programme to examine the role of quiescent galaxies at
z=2--2.5, we observed the field surrounding the radio loud quasar
1550-2655. This object lies at a redshift of 2.15 and shows signs of
associated Ly$\alpha$ absorption (Jauncey et al., 1984). Its spectrum
also contains a CIV absorber at z=2.09. Observations were made at the
3.5m ESO NTT using the SUSI optical imager for R band observations,
and the SOFI infrared imager for the K band observations. We acquired
a total of 4800s of integration time at R and 3600s at K. Data
reduction used standard IRAF and Eclipse procedures. The
observations were flux calibrated using the faint IR standards P499E
and S875C for K band, and the Landolt standard (Landolt, 1992)
PG1633+099C for R. Galactic extinction was also corrected.

After data reduction and flux calibration, the SUSI image, which has a
resolution of 0.13''/pixel, was rebinned to match the 0.292''/pixel
SOFI resolution, and the images were aligned. The final image was 67
by 88 arcseconds in size. The main limiting factor on this size was
the small SUSI field of view and the dithering scheme used for the
optical observations. We then used SExtractor to select objects
detected at K and to extract their photometric properties in matched
apertures in the two passbands. This matched catalogue can then easily
be searched for objects with specific colour criteria.  We detected a
total of 75 objects in K down to a limiting magnitude of $\sim20.5$.

We then searched the catalogue for candidate red objects, with
R-K$>$4.5.  We found five such objects in the catalogue, details of
which are given in Table 1. Their positions are also shown in Figure 1,
which shows both the R and K band images of the quasar field.

\subsection{A Red Quasar Companion}

Comparison of the R and K images of the quasar itself shows what would
appear to be a red companion object - apparent as an extension in K,
but absent in the R band image. The reality of this object was
investigated by subtracting off the unresolved quasar contribution.
This was achieved by selecting a star, with no close companions, in
the observed field and using this as a PSF model. The central value of
the PSF image was scaled to match that in the quasar image, and then
the two images were subtracted. The companion was clearly visible in
the K-band PSF subtracted image, but was entirely absent in the R-band
PSF subtracted image. The companion is marginally resolved, having a
size of roughly 1.5 x 1 arcseconds, and is situated $\sim$2 arcseconds
from the quasar.  R and K magnitudes were extracted from the PSF
subtracted image, indicating that the quasar companion is also
red. Its details are included in Table 1.

\vspace{1cm}
\begin{table}
\begin{center}
\begin{tabular}{crrr} \hline
Object No.&Kmag&Rmag&R-K\\ \hline
R1&20.2$\pm$0.1&$>$24.8&$>$4.6\\
R2&20.2$\pm$0.1&$>$25.3&$>$5.1\\
R3&20.4$\pm$0.1&$>$25.3&$>$4.9\\
R4&20.3$\pm$0.1&$>$24.9&$>$4.6\\
R5&20.3$\pm$0.1&$>$25.0&$>$4.7\\
C1&19.5$\pm$0.3&$>$24.5&$>$5.0\\
\end{tabular}
\caption{Properties of red companions to the RLQ 1550-2655}
All limits given are 3$\sigma$. The quasar companion is object C1.
\end{center}
\end{table}

\begin{figure}
\plottwo{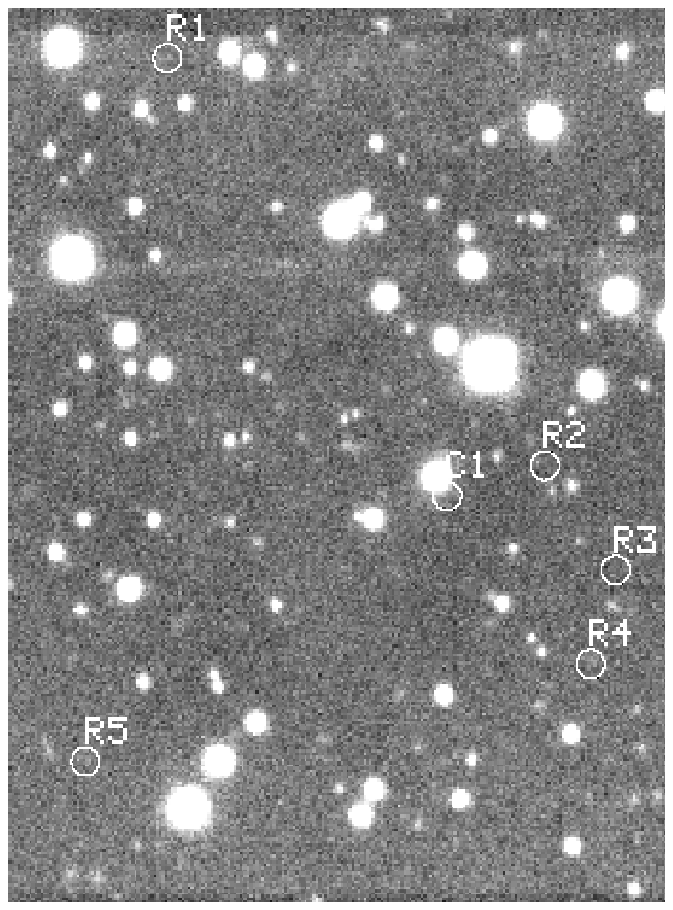}{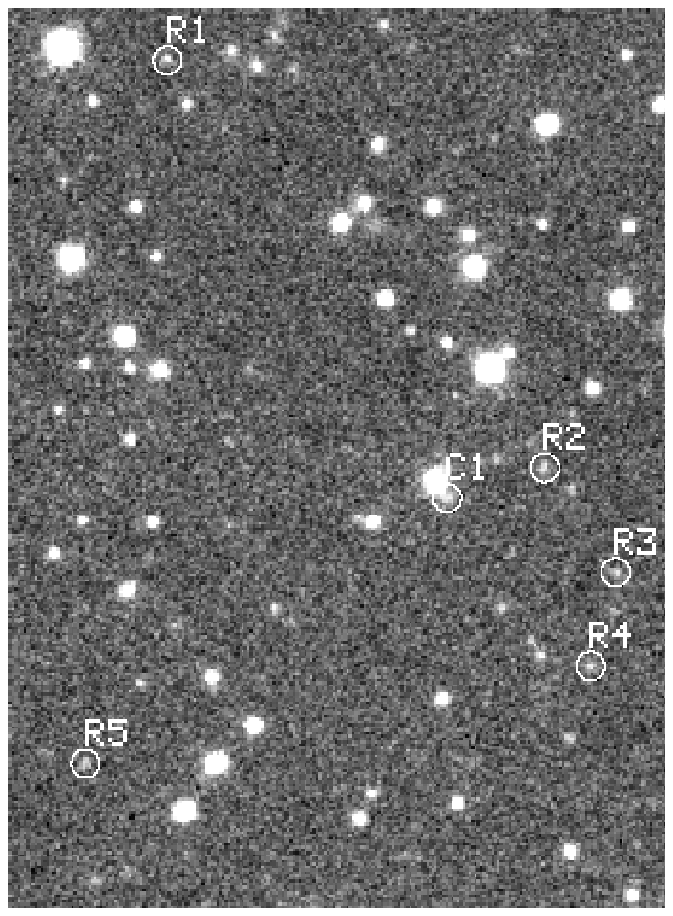}
\caption{R (left) and K band (right) images of the quasar field.}
The red associates are indicated in each image. Note their clear detections
in the K image, and their non-detections in the deeper R band image.
\end{figure}

\section{Discussion}

Until we can obtain spectroscopy for these objects, it is difficult to
assess their importance or role in this system. There are three possible
origins for these red objects: (1) they are associated with the quasar,
lying at z=2.15; (2) they are associated with the CIV absorber at
z=2.09; (3) they are foreground objects, unrelated to the quasar or CIV
system. We will assess each of these alternatives in turn.

\subsection{Association with the Quasar}

The density of red objects in this field is surprisingly high -
$\sim$4 arcmin$^{-2}$ as opposed to the supposed global value of
$\sim$1 arcmin$^{-2}$ (Cohen et al. 1998). The density of red objects
near to the quasar is significantly higher, with three of the red
objects as well as the companion lying in a 0.25 arcmin$^{-2}$ region
near the quasar. This is certainly suggestive that there is some
connection between the red objects and the AGN (or CIV system).  Other
studies have found a similar connection between red objects and AGN,
especially radio loud AGN. Perhaps the largest survey to date is that
of Hall et al. (1999) who obtained images of 31 z=1 -- 2 radio loud
quasars and found a significant excess of red galaxies around them.
Interestingly they found two radial dependencies for this excess, one
of which lies close to the quasar ($<$40'') and another more distant
(40''--100''). This is perhaps reflected in the present study, with
two red objects in the second class, further from the quasar, and the
rest, including the close companion, within 40''. In this
interpretation, the close companion would be $\sim$26kpc from the
quasar, and would be about 20x15 kpc in size. Hydrogen in this object
might be responsible for the associated absorption seen in the quasar
spectrum.

If we take the redshift of the companion galaxies to be the same as
the quasar, then the implied absolute magnitudes would be quite high,
$\sim$-27, ie. about 5 L$^{*}$ (Mobasher et al., 1993). This is not
without precedent. Francis et al. (1997) uncovered a group of
similarly red objects at z=2.38 associated with a cluster of quasar
absorption line systems. K magnitudes for these objects are similar
to, if not brighter then, the objects discussed here, implying still
greater absolute magnitudes. It is interesting to note that the
Francis et al. objects are all Ly$\alpha$ emitters. If the present red
objects have similar properties, then such line emission would make
redshift determination much easier.

\subsection{Association with the CIV Absorber}

Many of the same comments regarding direct association with the quasar
can be made regarding association with the CIV absorber: there is an
unusually high density of red objects in this region suggesting some
connection between them. Of particular interest here is the closeness
of the quasar companion to the quasar - only $\sim$2'' away, or 26kpc
at the redshift of the CIV absorber, and with a size similar to that
given above. To date little is known about the nature of CIV
absorbers, so the possible identification of the galaxy responsible
for one is rather interesting. Previous work looking for emission
lines from objects associated with damped and CIV absorbers (Mannucci
et al., 1998) suggests that galaxies are more likely to cluster with
absorbers than with quasars. If correct, this would suggest that the
objects found here are more likely to be associated with the absorber
than the quasar.

There has been a long and largely unsuccessful history of searches for
emission from absorption line systems in quasars at large
redshift. These have largely concentrated on emission lines, whether
Ly$\alpha$ (eg. Leibundgut \& Robertson, 1999), H$\alpha$, (eg. Bunker
et al., 1999), or others, though work in the infrared contiunuum has perhaps
shown greater success (see eg. Aragon-Salamanca et al., 1996,
1994). If the quasar companion in the present study is indeed
responsible for the CIV absorption, then we may have an explanation
for the failures.  The object is both red and quite close to the
quasar.  Detection of the companion would require both good seeing
(conditions for our own observations were sub-arcsecond) and
observations in the near-IR as well as the ability to subtract off the
quasar contribution. Sensitive infrared detectors have only recently
become available at most observatories, while subarcsecond seeing is
only rarely achieved. We might thus have been lucky in being able to
detect the companion. New instruments, such as UFTI at UKIRT, which
combines adaptive optics correction (regularly 0.5'') with a superb
infrared imager, can regularly make such observations. This will
hopefully allow us to make significant advances in our understanding
of high redshift absorption line systems.

\subsection{Foreground Contaminants}

The possibility that the red objects are at an entirely different
redshift to the quasar and absorber must still be considered while we
do not have confirming redshift spectra. In this context it is
salutary to note the lesson of the first VRO discovered (Hu \&
Ridgeway, 1994), known as HR10. This was found in the field of a
z=3.79 quasar, but was later shown to have a redshift of 1.44 (Graham
\& Dey, 1996).
However, the number density of red objects in their field was 0.9
arcmin$^{-2}$ which matches the field density of red objects discussed
by Cohen et al. (1998), and is lower than that found here.

\section{Conclusions}

At present there are several deficiencies in our data. Firstly we have
only obtained limits on the objects R band magnitudes. We must detect
them and measure, rather than limit, their R band magnitudes before we
can properly determine their colours. Secondly we must obtain spectra
for the objects so that we can actually determine, rather than
speculate, on their redshift. However, the results presented here
suggest that a larger survey of quasar environments, both with and
without absorbers, using infrared imagers with adaptive optics
correction might shed new light on galaxy populations at large redshift.

\acknowledgments
This paper is based on observations made at the European Southern
Observatory, Chile. It is a pleasure to thank Nick Devillard for his
excellent Eclipse data reduction pipeline, and E. Bertin for
SExtractor. I would like to thank Amanda Baker for useful
discussions. This work was support in part by an ESO fellowship, EU
TMR Network programme FMRX-CT96-0068 and by a PPARC postdoctoral
grant.

\end{document}